\documentclass[aps,pre,10pt,showpacs,floats,twocolumn,floats,superscriptaddress,floatfix]{revtex4}

\usepackage{graphicx}
\usepackage{bm}
\usepackage{amssymb}
\usepackage{color}
\usepackage{multirow}
\usepackage{amsmath}    
\usepackage{epsfig}
\usepackage{subfigure}  
\usepackage{hyperref}   
\usepackage{bm}
\usepackage{amssymb}
\usepackage{booktabs}
\usepackage{nicefrac}
\usepackage{booktabs}

\usepackage{float}
\usepackage[section]{placeins}
\graphicspath{{./}}

\definecolor{ForestGreen}{rgb}{0.1,0.5,0.1}
\definecolor{blau}{rgb}{0.1,0.2,0.8}

\usepackage{color}

\newcommand\RO{\textcolor{black}}
\newcommand\RT{\textcolor{black}}

\newcommand{\be}{\begin{equation}}
\newcommand{\ee}{\end{equation}}

\def\p{\partial}
\newcommand{\oPi}{\overline{{\Pi}}^\Delta}
\newcommand{\Pileo}{{\overline{\Pi}}^\Delta_{L}}

\newcommand{\oPis}{\overline{{\Pi}}^{\text{S}}_{\text{LES}}}
\newcommand{\oPie}{\overline{{\Pi}}^{\text{E}}_{\text{LES}}}
\newcommand{\oPia}{\overline{{\Pi}}^{\text{E}}_{\text{LBM}}}

\newcommand{\bk}{\bm{k}}
\newcommand{\bc}{\bm{c}}

\newcommand{\bx}{\bm{x}}
\newcommand{\by}{\bm{y}}
\newcommand{\br}{\bm{r}}
\newcommand{\bu}{\bm{u}}

\newcommand{\bhu}{\boldsymbol{\hat{u}}}

\newcommand{\ou}{\overline{ u}}

\newcommand{\obu}{\overline{\boldsymbol{u}}}
\newcommand{\oP}{\overline{{ p}}}
\newcommand{\otau}{\overline{{\tau}}^\Delta}
\newcommand{\otaus}{\overline{{\tau}}^{\text{S}}}
\newcommand{\otaue}{\overline{{\tau}}^{\text{E}}}
\newcommand{\otaua}{\overline{{\tau}}^{\alpha}}

\newcommand{\romeaddress}{Department of Physics and INFN, University of
Rome Tor Vergata, via della Ricerca Scientifica 1, 00133, Rome, Italy.}

\newcommand{\wuppertaladdress}{Chair of Applied Mathematics and Numerical Analysis, Bergische Universit\"{a}t Wuppertal, Gau\ss{}strasse 20, 42119 Wuppertal, Germany}

\begin{document}

\title{Inertial range statistics of the Entropic Lattice Boltzmann in 3D turbulence}

\author{Michele Buzzicotti}
\affiliation{\romeaddress}
\author{Guillaume Tauzin}
\affiliation{\romeaddress}
\affiliation{\wuppertaladdress}

\begin{abstract}
We present a quantitative analysis of the inertial range statistics produced by Entropic Lattice Boltzmann Method (ELBM) in the context of 3d homogeneous and isotropic turbulence. ELBM is a promising mesoscopic model particularly interesting for the study of fully developed turbulent flows because of its intrinsic scalability and its unconditional stability. 
In the hydrodynamic limit, the ELBM is equivalent to the Navier-Stokes equations with an extra eddy viscosity term~\cite{Malaspinas2008}. From this macroscopic formulation, we have derived a new hydrodynamical model that can be implemented as a Large-Eddy Simulation (LES) closure. This model is not positive definite, hence, able to reproduce backscatter events of energy transferred from the subgrid to the resolved scales.
A statistical comparison of both mesoscopic and macroscopic entropic models based on the ELBM approach is presented and validated against fully resolved Direct Numerical Simulations (DNS). Besides, we provide a second comparison of the ELBM with respect to the well known Smagorinsky closure. We found that ELBM is able to extend the energy spectrum scaling range preserving at the same time the simulation stability. Concerning the statistics of higher order, inertial range observables, ELBM accuracy is shown to be comparable with other approaches such as Smagorinsky model.

\end{abstract}

\pacs{}
\maketitle

\section{Introduction}
\label{sec:1}

Turbulence is common in nature and its unpredictable behavior has fundamental consequences on the understanding and control of various systems, from smaller engineering devices~\cite{dumitrescu2004rotational,Davidson2015,Pope2001}, up to the larger scales geophysical and astrophysical flows~\cite{gill2016atm,alexakis2018,barnes2001}.
Turbulent flows are described by the Navier-Stokes equations (NSE),
\begin{equation}
\begin{cases}
    \partial_t \bu + \bu \cdot \bm{\nabla} \bu = - \bm{\nabla} p + \nu \nabla^2 \bu + \bm{f} \\
    \bm{\nabla}\cdot\bu =0
\end{cases}
    \label{eq:nse}
\end{equation}
which give the evolution of the incompressible velocity field $\bu(\bx,t)$, with kinematic viscosity $\nu$, subject to a pressure field $p$ and to an external forcing $\bm{f}$.
However, even though \RT{the equations of motion are known} since almost two hundred years a direct analytical approach remains elusive~\cite{Frisch1995}. To overcome mathematical difficulties, scientists, helped by the exponential growth of the computational power, have tried to search for approximate solutions using numerical algorithms~\cite{fox2003computational,Pope2001,ishihara2009study}.
Unfortunately, also in this direction not every effort were successful. Indeed, the NSE have a very rich non-linear dynamics, where a large range of scales from the domain size up to the small scales fluctuations, are coupled together. This results in a very high dimensional problem, with the dimensionality proportional to the range of active scales~\cite{Frisch1995}, and with highly intermittent statistics dominated by the presence of extreme and rare fluctuations~\cite{yeung2015extreme,benzi2010inertial,iyer2017reynolds}. As a consequence, no matter how powerful new-supercomputers are, numerical algorithms cannot handle all the degrees of freedom involved in the dynamics~\cite{Davidson2015}. The way out is to introduce a scale separation and compute only the dynamics of degrees of freedom belonging to a subset of scales while neglecting the other ones \cite{Meneveau2000,filippova2001multiscale,Dong2008a,Dong2008b}.
However due to the non-linearity of NSE there is never a real scale separation in the equations of motion~\cite{bohr2005dynamical,Frisch1995}, and the small-scale effects on the scales of interest need to be compensated by the introduction of a model. 
In other words, the benefit of multi-scale modeling is to achieve a scale separation, and the main challenge is to find a ``closure'', which guaranties the numerical stability being at the same time the most accurate as possible in reproducing the coupling of the missing scales on the resolved ones. This is the principle behind the celebrated Large-Eddy simulations (LES), which actually solve the flow only on a subset of ``large'' scales by filtering each term of the NSE and replacing with a closure the non-linear coupling term between the resolved and the sub-grid scales (SGS) \cite{Pope2001,Lesieur2005}. One of the most important differences between the real coupling term coming from the filtered NSE and the common LES closures used in the literature is for the latter to be purely dissipative to ensure the simulation stability, see~\cite{Meneveau2000}. As a consequence, it is impossible for the closures to reproduce the backscatter events of energy going from the SGS to the large scales, with important consequences on the statistics of the resolved velocity field.
Another possible numerical approach, who has gained particular popularity, consists in solving the flow's macroscopic hydrodynamical properties as an approximation of its mesoscopic behaviour~\cite{de2013non}. The Lattice Boltzmann Method (LBM) falls into that category \cite{benzi1992lattice,Succi2001}. In LBM, the flow is simulated by evolving the Boltzmann equation for the single phase density function, $f(\bx,t)$. The idea is to evolve the streaming and collision of particles distribution functions, where the possible velocities are restricted on a subset of discrete lattice directions~\cite{frisch1986lattice,Wolf-Gladrow2000}. It is crucial to choose the collision operator so that macroscopically, (in the limit of small Knudsen number), the dynamics described by the NSE is recovered~\cite{qian1995recent,chen1998lattice}. The most common collision operator is the Bhatnagar-Gross-Krook model (BGK), see~\cite{Bhatnagar1954}, corresponding to the relaxation towards an equilibrium distribution, $f^{eq}_i(\bx,t)$, taken to be a discrete Maxwellian, with a fixed relaxation time, $\tau_0$,
\begin{align}
\begin{split}
    f_i(\bx+\bc_i\Delta_t,t + \Delta_t) - f_i(\bx,t) &=\\
    =- \frac{1}{\tau_{\rm 0}} \left [ f_i(\bx,t) - f^{eq}_i(\bx,t) \right ] & + F_i.
\end{split}
\label{eq:LBM}
\end{align}
Here, $F_i$, is a term introduced to model a macroscopic external forcing \cite{Succi2001}, and $i = 0, .., q-1$, indexes the different velocity directions on the lattice. Eq.~\eqref{eq:LBM} is obtained discretizing the Boltzmann equation and selecting the lattice spacing, $\Delta \bx$, such as divided by the time step, $\Delta_t$, they are equal to the lattice velocity $\bm{c}_i = \Delta \bx /\Delta_t$. From those mesoscopic quantities, it is then possible to recover the macroscopic velocity and density fields by following the perturbative Chapman-Enskog expansion~\cite{Wolf-Gladrow2000}. It can be shown that evolving Eq.~\ref{eq:LBM} is equivalent, up to approximations, to evolving the weakly compressible NSE for a flow with a density $\rho(\bx,t)=\sum_{i=0}^{q-1} f_i(\bx,t)$, a velocity $\bu(\bx,t)=\sum_{i=0}^{q-1} f_i(\bx,t)\bc_i / \rho(\bx,t)$ and a viscosity directly related to the relaxation time \cite{Wolf-Gladrow2000},
\RT{\begin{equation}
\nu_0 = c_{s}^{2} \Delta_t \left(\tau_0-\frac{1}{2}\right)\text{,}
\label{eq:lbm_viscosity}
\end{equation}
}where, $c_s$, is the speed of sound, \RT{and $\tau_0$ is the adimensional relaxation time}. A numerical validation of the hydrodynamic recovery of BGK-LBM in the context of 2D homogeneous isotropic turbulence (HIT) was performed in \cite{Tauzin2018} showing good agreement with DNS either in decaying that in forced regimes. Although this method is adapted to describing various physics of multi-phase and flows with complex boundaries in a highly scalable fashion, the BGK-LBM model suffers of numerical instabilities when, $\tau_{\rm 0} \rightarrow \frac{1}{2}$, \textit{i.e.} $\nu_{\rm 0} \rightarrow 0$, which has made the study of turbulent flows highly prohibitive for this method~\cite{Tauzin2018}.
To push LBM towards more turbulent regimes a number of collision operators have been proposed, see~\cite{eggels1996direct,yu2006turbulent,sagaut2010toward,malaspinas2012consistent}. Here we focus on the Entropic LBM (ELBM)~\cite{Karlin1999,ansumali2002single}, which tackles the stability issue by equipping LBM with an H-theorem. To achieve this results the ELBM differs from BGK-LBM by two major aspects. First, the equilibrium distribution $\bm{f}^{eq}(\bx,t)$ is not anymore a discretization of the Maxwell-Boltzmann distribution, but it is calculated as the extremum of a discretized H-function defined as;
\begin{equation}
H[\bm{f}]=\sum_{i=0}^{q-1}f_i\log\left(\frac{f_i}{w_i}\right), \,\, \bm{f} = \left\{f_i\right\}_{i=0}^{q-1},
\label{eq:H_functional}
\end{equation}
where $w_i$ are the weights associated to each lattice direction, under the constraints of mass and momentum conservation, see~\cite{Ansumali2003}. The second difference in ELBM, is that the relaxation time is not a constant anymore but is modified at every time step in order to enforce the non-monotonicity of H after the collision. This results in an apparent unconditional stability as $\nu_0 \rightarrow 0$~\cite{Karlin2015}. It follows that ELBM evolution equations are,
\begin{align}
\begin{split}
    f_i(x+c_i\Delta_t,t + \Delta_t) - f_i({\bm x},t) &=\\
    =-\alpha({\bm x},t)\beta [f_i({\bm x},t) &-f_i^{eq}({\bm x},t) ],
\label{eq:ELBM_eq}
\end{split}
\end{align}
where \RT{$\beta=1/(2\tau_0)$ is constant}, while the new relaxation time, \RT{$\tau_{\text{eff}}(\bx,t) = 1/(\alpha(\bx,t)\beta) $}, fluctuates in time and space through the definition of an entropic parameter $\alpha(\bx,t)$. \RO{More recently the ELBM method has been extended to a family of multi-relaxation time (MRT) lattice Boltzmann models~\cite{dorschner2016entropic,dorschner2017transitional,dorschner2018fluid}.} Note that $\alpha$ can be computed as the solution of the entropic equation $H[\bm{f}] = H[\bm{f}-\alpha\bm{f}^{neq}]$ which represents the maximum H-function variation due to a collision, with $\bm{f}^{neq}=\bm{f}-\bm{f}^{eq}$.
\RO{Following this approach the computation of $\alpha(\bx,t)$ can be performed via an expensive Newton-Raphson algorithm for every grid and at every time step of ELBM. To alleviate this problem, after the original ELBM formulation~\cite{Karlin1999}, a new version has been proposed where the computation of the entropic parameter is based on an analytical formulation derived as an first order expansion of the original model~\cite{karlin2014gibbs,bosch2015entropic}. However, to our knowledge, a study of high-order structure functions in the context of forced 3D HIT, has never been attempted before using the ELBM original model. In this regard, and also aiming to measure high-order, extremely sensitive statistical observables, we implemented the ELBM original formulation, relying on the least number of approximations, even though computationally more expensive.}
More details about ELBM will be given in section~\ref{sec:2}. Let us notice that BGK-LBM is recovered from Eq.~\eqref{eq:ELBM_eq} with $\alpha=2$ and the specific Maxwellian expression of $f_i^{eq}$.
It is important to stress that the bridge relation described in Eq.~\eqref{eq:lbm_viscosity} connecting viscosity and relaxation time still holds for fluctuating quantities, hence we can write,
\begin{align}
\begin{split}
\nu_{\text{eff}}\left(\alpha\right) &= c_{s}^{2} \Delta_t\left(\frac{1}{\alpha\beta}-\frac{1}{2}\right)=\\
 &=c_{s}^{2}\Delta_t\left(\frac{1}{2\beta}-\frac{1}{2}\right)+c_{s}^{2}\Delta_t\frac{2-\alpha}{2\alpha\beta}=\\
 &=\nu_0 + \delta \nu_\alpha,
\label{eq:entropic_viscosity}
\end{split}
\end{align}
where $\nu_0$ represents the constant kinematic viscosity and $\delta \nu_\alpha$ is the fluctuating term. Following this idea, it has been shown that ELBM is implicitly enforcing a SGS model of an eddy-viscosity type~\cite{Malaspinas2008,Karlin2015}. 
\RT{In particular, as initially done in~\cite{Malaspinas2008}, and then rederived in chapter 4 of~\cite{tauzin2019implicit}, performing a third-order Chapman-Enskog perturbative expansion in the limit of small Knudsen number (Kn), it is possible to obtain a macroscopic approximation of $\delta \nu_\alpha$, which can be written as,
\begin{equation}
 \delta \nu_\alpha^M = -c_s^2 \Delta t^2  \frac{S_{\ell j}S_{ij}S_{i\ell}}{S_{ij}S_{ij}},
\label{eq:ch5_nu_A}
\end{equation}}
where $S_{ij} = 1/2(\partial_j u_i+\partial_i u_j)$ is the strain rate tensor.
The entropic eddy viscosity in Eq.~\eqref{eq:ch5_nu_A} is particularly interesting because it is not positive definite and can reproduce events of energy backscatter, which is generally not the case among the other LES closures, i.e. see the Smagorinsky eddy viscosity~\cite{Smagorinsky1963}. 
After the introduction of the new LES model, see section~\ref{sec:2}, we compare it with standard Smagorinsky closure and with fully resolved data obtained from DNS. At the same time, we also present for the first time a quantitative investigation of the inertial range statistics provided by the mesoscopic ELBM approach in the context of 3d turbulence.
Results provide evidence that ELBM is a good approximation of the 3d flows up to turbulent regimes never reachable with the standard BGK-LBM. We found that ELBM guaranties the simulation stability producing a considerable extension of the inertial range scaling, i.e. the extension of the energy spectrum power law. Measuring statistical properties of higher order observables such as the structure functions, we found that ELBM is also able to reproduce qualitatively the intermittent features of real 3d turbulent flows with an accuracy comparable to the Smagorinsky model. To get a more accurate estimation of the anomalous exponents more refined models are required~\cite{biferale2019self}.\\

The paper is organized as follows. In section~\ref{sec:2} we introduce the details of the ELBM and the LES models considered in this work. In section ~\ref{sec:3} we present the set of simulations performed. In sect.~\ref{sec:4} we evaluate the quality of the two LES closures by comparing them with the real SGS energy transfer measured from fully resolved DNS. In sect.~\ref{sec:5} we focus on the intermittent properties by analysing high-order inertial range statistics~\cite{Arneodo2008,benzi2010inertial}. In sect.~\ref{sec:6} we discuss our conclusions.

\section{Turbulence Modelling}
\label{sec:2}

In this section we give a description of the SGS modelling approaches considered in this work. We start discussing the mesoscopic ELBM approach highlighting the differences with respect to the standard BGK-LBM. Following we discuss the new hydrodynamic LES closure inspired by the ELBM macroscopic approximation first derived in~\cite{Malaspinas2008}. In the end of this section we briefly recall the well known Smagorinsky model.

\subsection{Entropic Lattice Boltzmann Method}

Using the same formalism as in~\cite{Karlin1999}, the ELBM, Eq.~\eqref{eq:ELBM_eq}, can be rewritten as,
\begin{align}
\begin{split}
f_{i}(x+c_{i} &\Delta_t,t + \Delta_t) =\\
=&f_{i}({\bm x},t)-\alpha({\bm x},t)\beta\left(f_{i}({\bm x},t)-f_{i}^{eq}({\bm x},t)\right)\\
=&\left(1-\beta\right)\, f^{pre}_{i}({\bm x},t)+\beta\, f_{i}^{mir}({\bm x},t)\\
=& f_i^{post}(\bx,t),
\end{split}
\label{eq:ELBM_eq_2}
\end{align}
where the fluctuating relaxation time is $\tau_{\text{eff}}({\bm x},t) = 1/(\alpha({\bm x},t) \beta)$, with  $\beta=1/(2\tau_0)$ and where $\alpha(\bx,t)$ is the time and space dependent, locally-calculated, entropic parameter. The post-collision distribution, $\bm{f}^{post}(\beta)$, can be understood as a convex combination between the pre-collision distribution, $\bm{f}^{pre} = \bm{f}$, and the so-called mirror distribution, $\bm{f}^{mir}(\alpha)=\bm{f}^{pre} - \alpha \, \bm{f}^{neq}$, with $\bm{f}^{neq}=\bm{f}^{pre}-\bm{f}^{eq}$, the non-equilibrium part of $\bm{f}^{pre}$. This convex combination is parametrized by the parameter $\beta$ in the range $0 < \beta < 1$ for which we have $0.5 < \tau_0 < +\infty$. From the definition of the H-functional given in Eq.~\eqref{eq:H_functional} the discrete H-theorem can then be expressed as a \RT{the local decrease of the H-functional} between the pre-collision and post-collision distributions,
\begin{align}
\begin{split}
\Delta H &= H[\bm{f}^{post}] 
- H[\bm{f}^{pre}] \\
&= H[(1-\beta) \bm{f}^{pre} + \beta \bm{f}^{mir}(\alpha)] 
- H[\bm{f}] \leq 0,
\label{eq:H_inequality}
\end{split}
\end{align}
The equilibrium distribution function $\bm{f}^{eq}$ can be calculated as the extremum of the convex H-functional introduced in Eq.~\eqref{eq:H_functional}, which has an analytical solution for the D1Q3 lattice, whose tensorial product is solution for three D2Q9 and the D3Q27 lattice,
\begin{align}
\begin{split}
    f_i^{eq}&(\bx,t) =\\
    w_i \rho &\prod_{j=1}^d \left \{ \left ( 2 - \sqrt{1 + \frac{u_j^2}{c_s^2}} \right ) \left [ \frac{\frac{2 u_j}{\sqrt{3}c_s} + \sqrt{1+ \frac{u_j^2}{c_s^2}}}{1-\frac{u_j}{\sqrt{3}c_s}} \right ]^{\frac{c_{ij}}{\sqrt{3}c_s}} \right \},
    \label{eq:Feq_entropic}
\end{split}
\end{align}
where $d$ is the dimension of the DdQq lattice, $c_s$ is the speed of sound, $w_i$ are the weights associated with each lattice direction and $u_j(\bx,t)=\sum_{i=0}^{q-1} f_i(\bx,t)c_{ij} / \rho(\bx,t)$ is the flow macroscopic velocity. It is important to remark that the first three moments of the entropic equilibrium distribution Eq.~\eqref{eq:Feq_entropic} are exactly the same as the one coming from the 3\textsuperscript{rd} order Hermite polynomial expansion of the Maxwell-Boltzmann equilibrium distribution, namely,
\begin{align}
\begin{split}
f^{\rm{eq}}_{i} (\rho, {\bm u}) = w_{i} \rho &\bigg( 1 +\frac{{\bm u}\cdot{\bm c}_{i}}{c_s^{2}} +\frac{{\bm u}{\bm u}:{\bm c}_{i}{\bm c}_{i} - c_s^{2}|{\bm u}|^2}{2 c_s^{4}} \\
&+\frac{{\bm u}{\bm u}{\bm u} :\!\cdot\, {\bm c}_{i}{\bm c}_{i}{\bm c}_{i} - 3c_s^2|{\bm u}|^2{\bm u}\cdot{\bm c}_{i}}{6c_s^6} \bigg),
\label{eq:feq_Ma3}
\end{split}
\end{align}
\RT{therefore, it allows the recovery of the same athermal weakly compressible NSE as in the case of BGK-LBM~\cite{Ansumali2003}. This recovery, obtained by performing a Chapman-Enskog expansion at the second order in Kn, is also valid for ELBM as fluctuation of $\alpha$ around $2$ leads to higher-order terms in Kn, absorbed in $\mathcal{O}(\text{Kn}^2)$~\cite{tauzin2019implicit}.}
In this work, following the approach used in~\cite{Karlin1999}, we calculate $\alpha(\bx, t)$ as the solution of,
\begin{equation}
H[\bm{f}^{pre}]=H[\bm{f}^{mir}],
\label{eq:entropic_step}
\end{equation}
which can be estimated via the popular Newton-Raphson algorithm. In this way, $f_i^{post}$ being a convex combination between two distributions, $f_i^{pre}$ and $f_i^{mir}$ of equal H-value, and being H at the same time a convex functional, the monotonic decrease of the H is ensured. 
\RT{Let us stress that, as it was shown in~\cite{tauzin2019implicit}, the ELBM equation cannot be considered macroscopically as an approximation to the weakly compressible NSE with the addiction of a sole eddy viscosity term of the form of Eq.~\eqref{eq:ch5_nu_A}. Indeed, this term appears in a macroscopic equation of motion that requires a Chapman-Enskog expansion of third order in the Knudsen number, while the NSE are recovered at the second order. As a consequence a number of extra third-order terms are part of the implicit ELBM SGS model.} This makes the actual ELBM closure even more complex than a simple eddy viscosity, and in principle, able to outperform standard methods. 
On top of this, as already discussed in the introduction, the macroscopic approximation of the ELBM eddy viscosity, Eq.~\eqref{eq:ch5_nu_A}, has itself a very interesting formulation, being similar to a Smagorinsky eddy viscosity~\cite{Smagorinsky1963}, but being not positive-definite and therefore allows events of energy backscatter, \textit{i.e.} energy transfer from the unresolved to the resolved scales. Indeed, while energy in 3d turbulence is on average cascading from the large towards the small scales, in real flows there are local events of energy going backward with non-trivial implications on the statistical properties of the resolved scales.

\subsection{Large-Eddy Simulations}

The LES governing equations can be directly derived by filtering each term of the incompressible NSE, see~\cite{Lesieur2005,Meneveau2000}. The filtering operation consist in a convolution between the full velocity field and a filter kernel. There are several choices that can be made for the filter kernel, see~\cite{Pope2001}, in this work, we consider a ``sharp spectral cutoff'' in Fourier space. This choice is convenient for two reasons, first, the sharp cutoff produces a clear separation between resolved and sub-grid scales, defined respectively as all scales above and below the cutoff wavenumber, $k_c$. Second, it is a Galerkin projector that produces the same results when operating multiple times on a field, which allows to have a clear scale separation also in a dynamical sense, namely it projects on the same support all terms (non-linear ones included) of the equations of motion, see~\cite{buzzicotti2018effect}.
In the following we briefly sketch the main operations required to arrive at the LES equations. Given a filter kernel $G_{\Delta}(\bx)$, the filtered velocity $\obu(\bx,t)$ can be defined by the following convolution operation, 
\begin{align}
\begin{split}
\obu(\bx,t) \equiv \int_\Omega  d\by \ G_\Delta(|\bx-\by|) \, \bu(\by,t) =\\
=\sum_{\bk \in \mathbb{Z}^3} \hat G_\Delta(|\bk|) \bhu(\bk,t) e^{i \bk \bx} \,
\end{split}
\end{align}
where $\hat G_\Delta$ is the Fourier transform of $G_\Delta$, and $\Delta \sim \pi/k_c$, is the filter cutoff scale, see \cite{Pope2001}. Applying the filtering operation to all terms in the Navier-Stokes equations we get,
\be
\p_t \obu  + \nabla \cdot(\overline{\obu \otimes \obu}) = -\nabla \oP -\nabla \cdot \otau(\bu,\bu)+ \nu \Delta \obu \ . 
\label{eq:Ples}
\ee
Here, we have introduced the SGS tensor, $\otau(\bu,\bu)$, defined as,
\be
\label{eq:tau_ples}
 \otau_{ij}(\bu,\bu) =  \overline{u_iu_j} - \overline{\ou_i\ou_j} ,
\ee
which is the only term of Eq.~\eqref{eq:Ples} that depends on SGS scales. Hence, it is the only term that needs to be replaced by a model to close the equations in terms of the resolved-scales dynamics. From $\otau(\bu,\bu)$, we can easily get the exact formulation of the SGS energy transfer $\oPi$, namely, the energy transfer across the filter scale produced by the real non-linear coupling in the NSE. To do so we need to multiply with a scalar product each term of Eq.~(\ref{eq:Ples}) and the velocity field to obtain the filtered energy balance equations;
\be
\label{eq:sg-eneP-leo}
\frac{1}{2}\p_t(\ou_i \ou_i) + \p_j A_{ij} + \Pileo =  -\oPi . 
\ee
The terms on the lhs of Eq.~\eqref{eq:sg-eneP-leo} are defined respectively as $\partial_j A_{j} = \partial_j \ou_i ( \overline{\ou_i \ou_j} + \overline{p} \delta_{ij}+ \otau_{ij} - \frac{1}{2} \ou_i\ou_j)$ and $\Pileo = - \partial_j \ou_i \left ( \overline{\ou_i\ou_j}-\ou_i\ou_j \right )$. 
\begin{table*}
 \begin{center}
 \begin{tabular}{cccclcll}
  ID & & Eddy viscosities & & Stress tensors & & & Energy transfers   \\
  \hline
 DNS& & --- & &$\otau_{ij} =\overline{u_iu_j} - \overline{\ou_i\ou_j}$ & & & $\oPi = -\partial_j \ou_i \,\, \otau_{ij}(\bu,\bu)$ \\
 S-LES   & & $\nu^{S}_e = (C_S \Delta)^2  \sqrt{2 S_{ij}S_{ij}} $ &  &$\otaus_{ij}= -2 \nu_e^{S} \bar{S}_{ij} $ & & & $\oPis = -\otaus_{ij} \overline{S}_{ij}$ \\
 E-LES   & & $\nu^E_e = (C_E \Delta)^2  \frac{S_{\ell j}S_{ji}S_{i\ell}}{S_{ij}S_{ij}} $ &  &$\otaue_{ij}= -2 \nu_e^{E} \bar{S}_{ij} $ & & & $\oPie = -\otaue_{ij} \overline{S}_{ij}$ \\ 
 ELBM    & & $\delta \nu_\alpha = c_{s}^{2}\frac{2-\alpha}{2\alpha\beta}\Delta_t $ &  &$\otaua_{ij} = -2 \delta \nu_\alpha \bar{S}_{ij} $ & & & $\oPia = -\otaua_{ij} \overline{S}_{ij}$ \\
  \hline
  \end{tabular}
  \end{center}
 \caption{Summary of definitions of eddy viscosities, sub-grid scales stresses and energy transfers. The ID column indicates the names of the corresponding set of simulations, see sect.~\ref{sec:3}, in particular S-LES and E-LES correspond to the hydrodynamical LES respectively with Smagorinsky and macroscopic ELBM model, while with ELBM we refer to the SGS energy transfer measured from the mesoscopic quantities. 
 }
 \label{tbl:definitions}
\end{table*}
As shown in \cite{leonard1974energy,buzzicotti2018effect}, to get the correct contribution to the energy transfer, it is important to distinguish between $\Pileo$ and the actual SGS energy transfer $\oPi$ because the former depends only on resolved-scales quantities and does not contribute to the mean energy flux across the cutoff scale. On the other hand, 
\be
\label{eq:SGS_Pi_ples}
 \oPi= -\partial_j \ou_i \,\, \otau_{ij}(\bu,\bu) = -\partial_j \ou_i \left ( \overline{u_iu_j} -  \overline{\ou_i\ou_j} \right ), 
\ee
is the flux which depends on both the SGS and the resolved scales. In this work, as already mentioned, we consider as possible closure the Smagorinsky LES model (referred to as S-LES in the sequel), $\otaus_{ij}(\obu,\obu)$, which aims to model the deviatoric part of the stress tensor, $\otau_{ij} - \frac{1}{3} \otau_{kk}\delta_{ij} \,\, \rightarrow \,\, \otaus_{ij}$, as follows,
\begin{equation}
\otaus_{ij} = -2 \nu_e^{S}(\bx,t) \bar{S}_{ij}, \quad \nu^{S}_e = (C_S \Delta)^2  \sqrt{2 S_{ij}S_{ij}}
\label{eq:smag}
\end{equation}
where, $\bar{S}_{ij} = 1/2 (\p_j \ou_i + \p_i \ou_j)$, is the resolved scales strain-rate tensor, $\nu_{e}^S$ is the Smagorinsky eddy viscosity depending on the filter cutoff scale $\Delta$ and the non-dimensional factor $C_S$. From the definition of the macroscopic approximation of the ELBM eddy viscosity in Eq.~\eqref{eq:ch5_nu_A}, we can now define the hydrodynamic ELBM-LES model (called E-LES in the sequel),
\begin{equation}
\otaue_{ij} = -2 \nu_e^{E}(\bx,t) \bar{S}_{ij}, \quad \nu^{E}_e = (C_E \Delta)^2 \frac{S_{\ell j}S_{ji}S_{i\ell}}{S_{ij}S_{ij}},
\label{eq:entropic}
\end{equation}
where $C_E$ is the entropic dimensionless coefficient. \RT{Comparing the definition of $\nu^{E}_e$ with $\delta \nu_\alpha^M$, in Eq.~\eqref{eq:ch5_nu_A}, we can see that they both have the same functional dependency on the strain-rate tensor, but different signs and multiplicative constants. In particular, the minus sign of the E-LES closure has been absorbed in the definition of $\otaue_{ij}$ to align with the Smagorinsky closure formulation.} Let us stress that the E-LES model has the same scaling as the Smagorinsky model, proportional to the strain rate, but it is not positive definite.
From the above definitions of the S-LES and E-LES models the two corresponding SGS energy transfers can be written as,
\begin{equation}
\oPis = -\otaus_{ij} \overline{S}_{ij}; \qquad \oPie = -\otaue_{ij} \overline{S}_{ij}.
\label{eq:SGS_pi_model}
\end{equation}
To compare the behaviour of the mesoscopic ELBM model with respect to the two hydrodynamical approaches just introduced, we can approximate the SGS energy transfer from the ELBM as, 
\begin{equation}
\oPia = - 2 \delta \nu_\alpha \overline{S}_{ij}\overline{S}_{ij},
\label{eq:SGS_pi_model}
\end{equation}
where $\oPia$ stands for ELBM SGS energy transfer and $\delta \nu_\alpha=c_{s}^{2}\Delta_t\frac{2-\alpha}{2\alpha\beta}$ is the mesoscopic fluctuating viscosity depending on $\alpha(\bx,t)$. The strain rate tensor can be measured from the ELBM data after the calculation of the macroscopic velocity in terms of the mesoscopic ones, namely, $\bu(\bx,t)=\sum_{i=0}^{q-1} f_i(\bx,t)\bc_i / \rho(\bx,t)$. A summary of these SGS energy transfer definitions with their respective SGS tensors, eddy viscosities is given in table~\ref{tbl:definitions}. \RT{It is worth noting that the ELBM in the limit of small Knudsen numbers is not equivalent to the entropic LES. Indeed, as previously mentioned, the eddy viscosity term appears in the Chapman-Enskog expansion at the third order in Kn along with various extra terms that are not contained in the entropic LES formulation.}

\section{Numerical Simulations}
\label{sec:3}

\begin{figure}[htp]
\centering
\includegraphics[width=7cm]{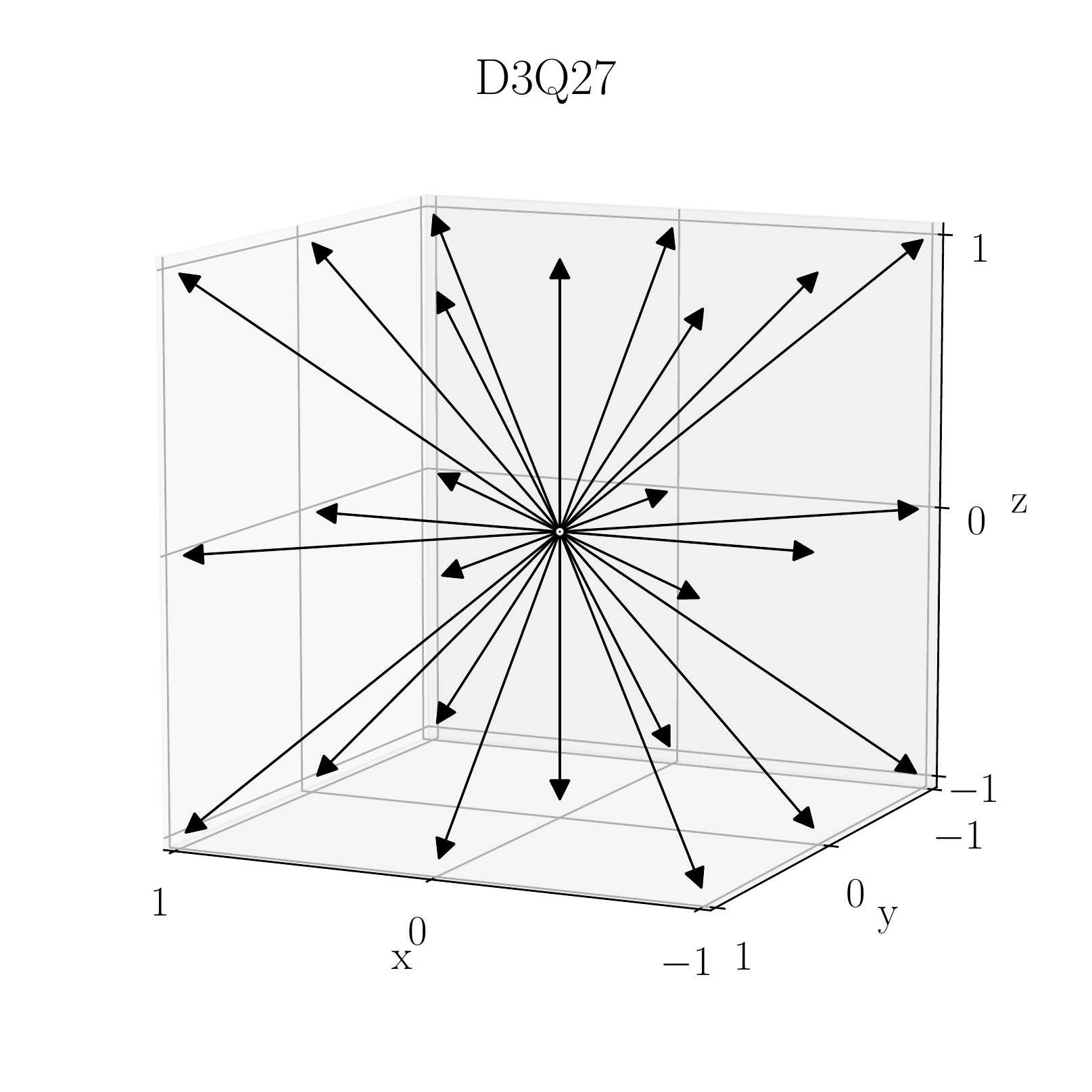}
\caption{Schematic representation of the D3Q27 lattice stencil used for the ELBM simulation.}
\label{fig:d3q27}
\end{figure}

\begin{figure}[htp]
\centering
\includegraphics[width=0.5\textwidth]{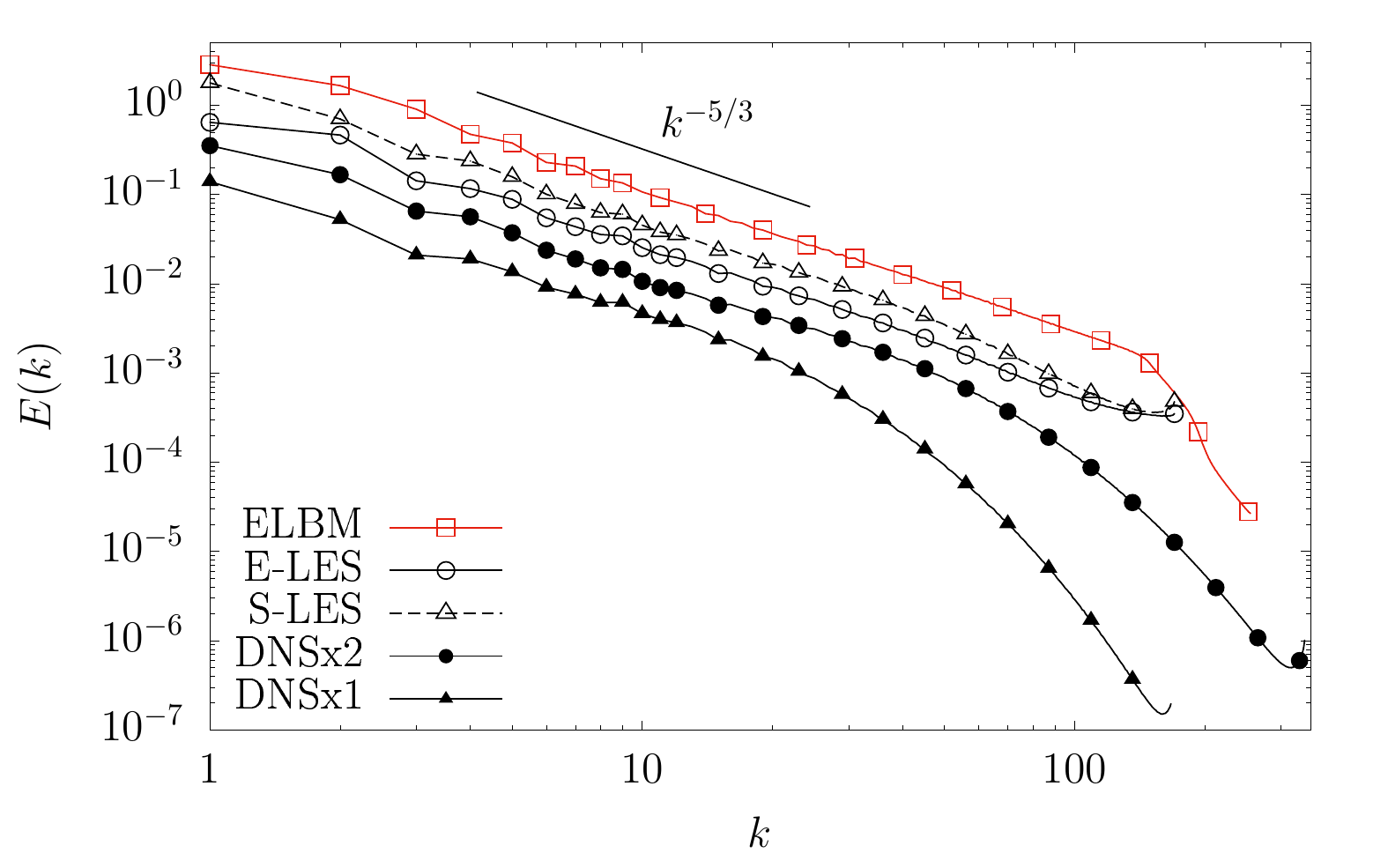}
\caption{Time-averaged spectra for the conducted simulations at $512^3$ grid points, measured from the mesoscopic ELBM simulation (empty squares, red color), the hydrodynamical LES with entropic model (E-LES, empty circles) and with Smagorinsky model (S-LES, empty triangles). The energy spectra from fully resolved DNS at $512^3$ (DNSx1) and $1024^3$ (DNSx2) are presented respectively with full triangles and full circles. The curves are shifted vertically for the sake of data presentation. The Kolmogorov predicted slope of $k^{-5/3}$ is given as a reference~\cite{Frisch1995}.}
\label{fig:spectra}
\end{figure}

All simulations performed in this work are intended to model HIT on a three dimensional domain with periodic boundary conditions. In the following we provide some details about the sets of simulations performed with the different modelling techniques.
Concerning the lattice Boltzmann simulation with entropic formulation of the relaxation time, ELBM, we have conducted a set of 3d simulations with a number of $512$ collocation points along each spatial direction. To reach stationarity the flow is forced at large scales, $1 \le |\bk|\le 2$ with a constant and isotropic forcing. \RT{More precisely we have used in all simulations the same forcing, defined in Fourier space with constant phases and amplitudes, added isotropically to all wave-vectors at large-scales.} To ensure incompressibility the forcing is projected on its solenoidal component. The ELBM simulation uses a lattice with 27 discrete velocities (see Fig.~\ref{fig:d3q27}), the D3Q27~\cite{Succi2001, Wolf-Gladrow2000, Kruger2017}. The spectral forcing is implemented using the exact-difference method forcing scheme~\cite{Kuperstokh2004} for a relaxation time $\tau_0 = 0.5003$ corresponding to $\beta \approx 0.9994$.
Considering the LES we have performed two sets of pseudo-spectral fully dealiased simulations on a domain $\Omega =[0,2\pi]^3$ with periodic boundary conditions both at the resolution of $512^3$ grid points. The first LES is equipped with the Smagorinsky model (S-LES), see Eq.~\eqref{eq:smag}, and a second with the macroscopic formulation of the entropic eddy viscosity, see Eq.~\eqref{eq:entropic}, (E-LES). \RT{As discussed above, }all simulations are forced with the same isotropic constant forcing mechanisms acting only on the larger system scales ($1 \le |\bk|\le 2$). In the expression of Smagorinsky eddy viscosity Eq.~\eqref{eq:smag}, we use the standard value of $C_S = 0.16$, while for the \RT{entropic eddy viscosity Eq.~\eqref{eq:entropic}, we use $C_E = 0.45$}, found to be optimal values for the best compromise between the maximization of inertial range extension and the minimization of spurious effects produced by the model~\cite{buzzicotti2020synch}. For both we have $\Delta = \pi/k_{max}\approx 0.0184$ with $k_{max}=171$, which comes from the $2/3$ rule for the dealiasing projection \cite{Patterson71}.
Additionally, as a reference, we have run a pseudo-spectral fully resolved DNS of the NSE with the same forcing scheme on the same 3d domain $\Omega =[0,2\pi]^3$, using a number of $512^3$ (DNSx1) and $1024^3$ (DNSx2) collocation points. The resolution in both DNS is kept such as $\eta_\alpha/dx \simeq 0.7$, where $dx$ is the grid spacing and $\eta_\alpha = (\nu^3/\epsilon)^{1/4}$ is the Kolmogorov microscale \cite{Borue95} with $\varepsilon$ denoting the mean energy dissipation rate.
In order to create ensembles of statistically independent data all the ELBM, LES and DNS are sampled on time intervals of one large-scale eddy turnover time after reaching a statistically stationary state. It is worth mentioning that both the ELBM and E-LES simulations remain stable even though their modeling terms are not purely dissipative being their eddy viscosities not positive definite.
\RT{In Fig.~\ref{fig:spectra} we show} the time-averaged energy spectra, $E(k)$, for all simulations. It is visible that all modelled simulations have an extended inertial range with respect to DNS at the same resolution (DNSx1) with an inertial range slope close to the Kolmogorov prediction of $k^{-5/3}$~\cite{Frisch1995}. Let us stress that the ELBM spectrum reaches a maximum wavenumber higher than the pseudo-spectral data, this is because in the ELBM simulations there is not any dealiasing operation. Anyway also in the ELBM case the spectrum loses the Kolmogorov inertial range scaling at wavenumber larger that the pseudo-spectral dealiasing cutoff.

\section{SGS energy transfer analysis}
\label{sec:4}

\begin{figure*}[htp]
\centering
  \includegraphics[height=4.8cm]{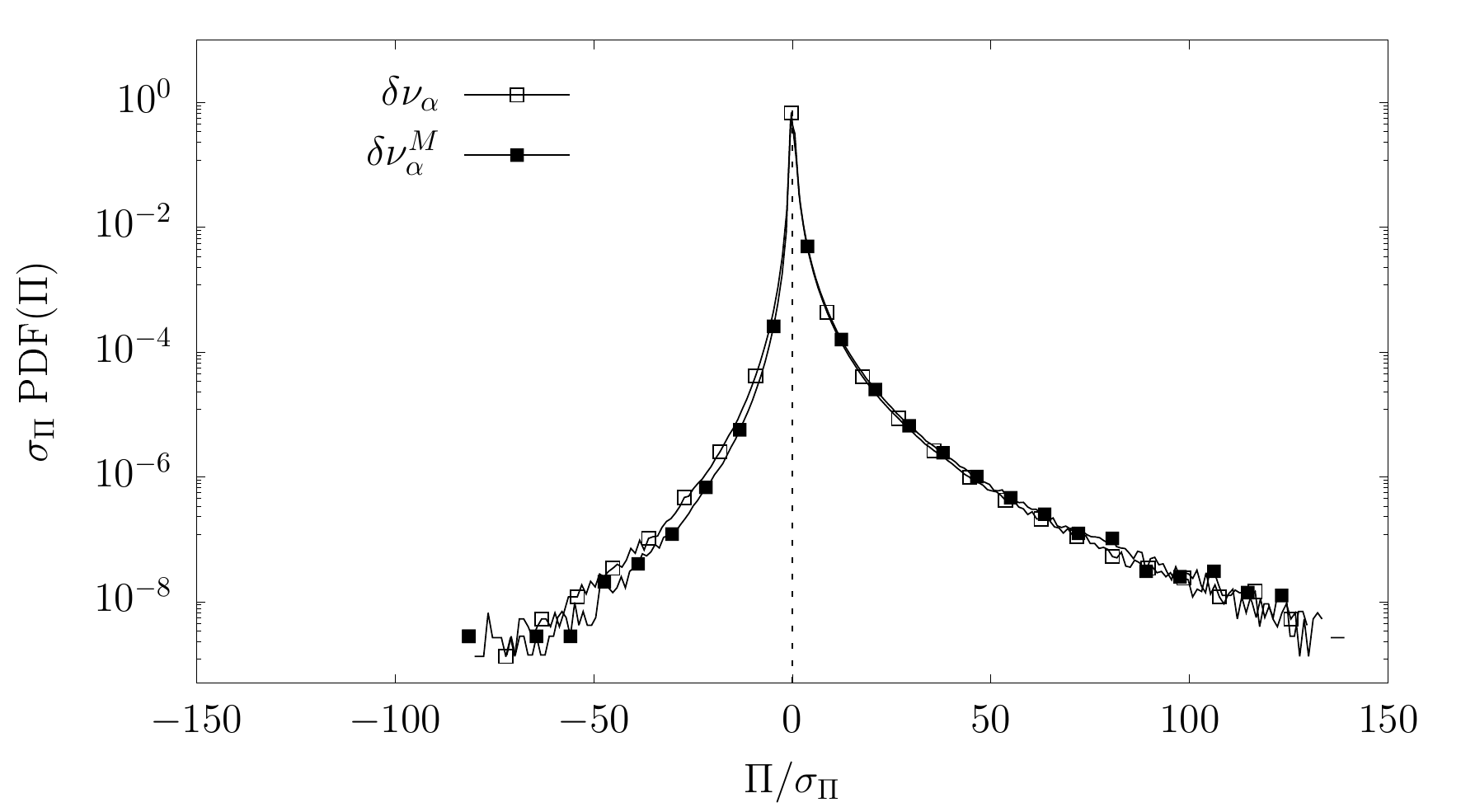}
  \includegraphics[width =9cm]{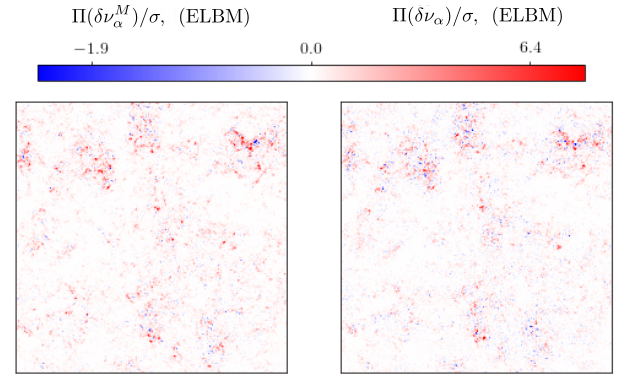}
\caption{Standardized PDFs of the ELBM SGS energy transfers measured from the correct fluctuating viscosity, $\delta \nu_\alpha$, and its approximated formulation, \RT{$\delta \nu_\alpha^M$} (left panel). Visualization of a plane of ELBM SGS energy transfer measured using the approximated (center panels) and correct definitions of eddy viscosity (right panel).}
\label{fig:compare_sgs}
\end{figure*}

\begin{figure}[htp]
  \centering
  \includegraphics[width=0.5\textwidth]{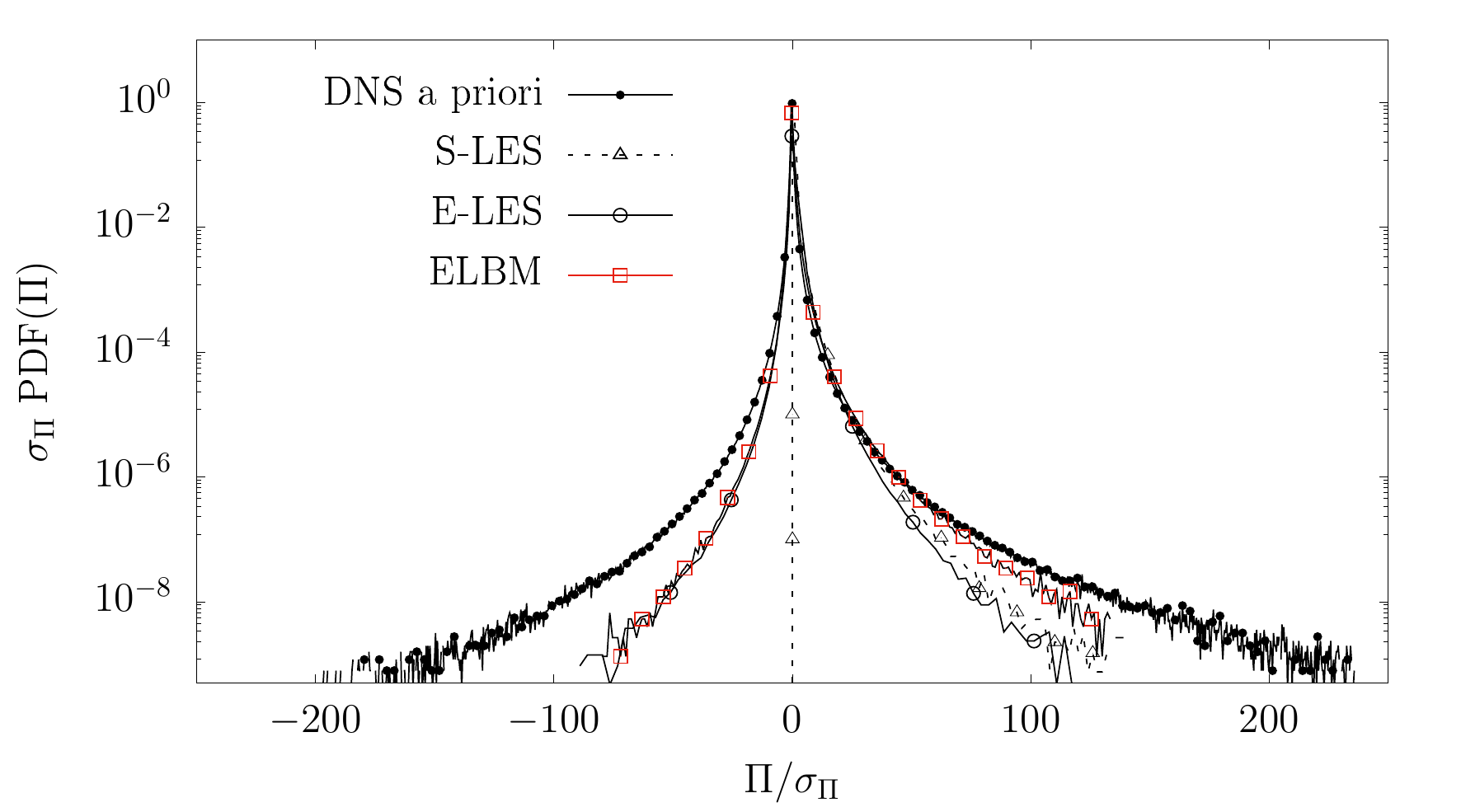}
  \caption{Standardized PDF of the SGS energy transfer measured from the a posteriori data obtained via the ELBM simulations (empty squares, red color), the LES with hydrodynamical entropic closure (E-LES, empty circles) and from LES with Smagorinsky model (S-LES, empty triangles). For comparison the PDF measured for the real SGS energy transfer measured a priori by filtering data from higher resolution simulations is presented (DNSx2, full circles).}
    \label{fig:pdf_sgs}
\end{figure}

In this section we provide a statistical comparison of the SGS energy transfers measured in the modelled simulations together with the original SGS transfer measured {\em a priori} from higher resolution DNS, see Table~\ref{tbl:definitions}.
Before staring the comparison between macroscopic and mesoscopic simulations, we analyse the quality of the approximation made in the Chapman-Enskog expansion to obtain the macroscopic formulation of the ELBM eddy viscosity \cite{Malaspinas2008}. 
\RT{In this direction, we have computed the SGS energy transfer defined in Eq.~\eqref{eq:SGS_pi_model} using the two different definitions of the fluctuating viscosity. Namely, either the correct definition, $\delta \nu_\alpha$, depending on the entropic parameter or its third order expansion in the limit of small Kn, $\delta \nu_\alpha^M$ see Eq.~\eqref{eq:ch5_nu_A}.} Their statistical comparison is shown in Fig.~\ref{fig:compare_sgs}, where on the left panel we show the probability density functions (PDF) measured from the ELBM SGS energy transfer using the two different formulations. Here we can see that the PDFs once re-scaled by their standard deviations have almost an identical shape. From the center and right panels of the same figure, we can qualitatively see two visualizations of the SGS energy transfers measured by selecting the same plane of the velocity field. From these visualizations we can appreciate that there is a very high spatial correlation between them. These results suggest that the approximation of neglecting the extra third order terms coming from Chapman-Enskog expansion is a good approximation of the ELBM eddy viscosity.
Let us now analyse the statistics of the SGS energy transfers, comparing them also with the statistics of the real SGS energy transfer, see Eq.~\eqref{eq:SGS_Pi_ples}, measured a priori from fully resolved DNSx2. To obtain the a priori $\oPi$ we filter the velocity field with a sharp projector in Fourier space with a cutoff at the maximum wavenumber allowed in the modelled simulations, which corresponds to the dealiasing cutoff ($k_{max} = 171$). As known, the presence of a forward energy cascade, as in 3d turbulence, reflects in a skewed PDF of the a priori $\oPi$, see~\cite{gotoh2005statistics,buzzicotti2018energy,buzzicotti2018effect}, see Fig.~\ref{fig:pdf_sgs}. Instead, the negative tail describes the presence of intense backscattering events with fluctuations up to two orders of magnitude larger than the standard deviation. The main remarkable difference between the different models considered here is that, as expected, the ELBM and E-LES produce backscatter events, while the Smagorinsky model is positive definite in its energy formulation and it produces a zero tail in the negative region. Let us notice that ELBM mesoscopic model shows qualitatively a better overlap with respect to the DNS data.

\section{Inertial range statistics}
\label{sec:5}
\begin{figure*}
  \centering
  \includegraphics[width=8.5cm]{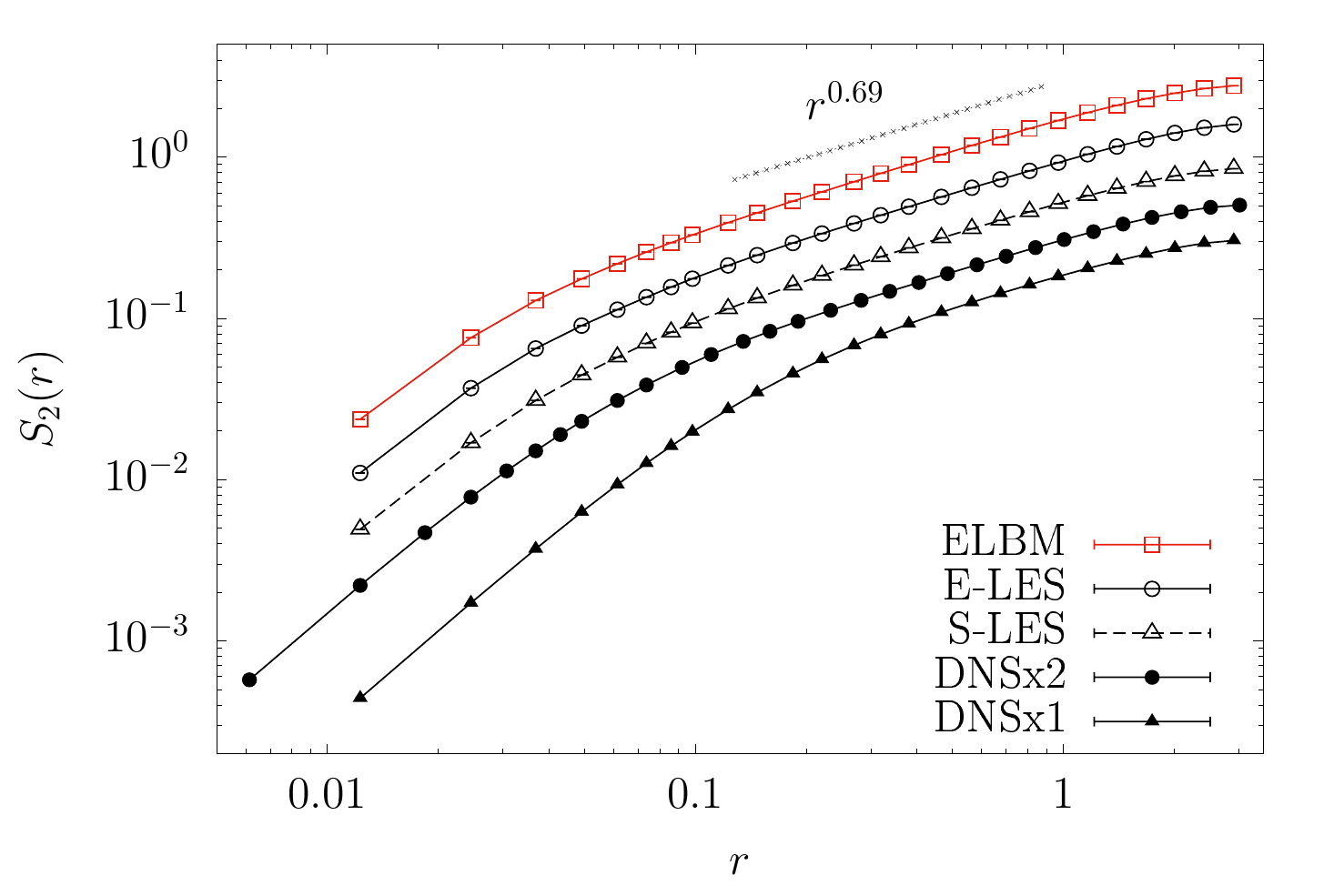}
  \includegraphics[width=8.5cm]{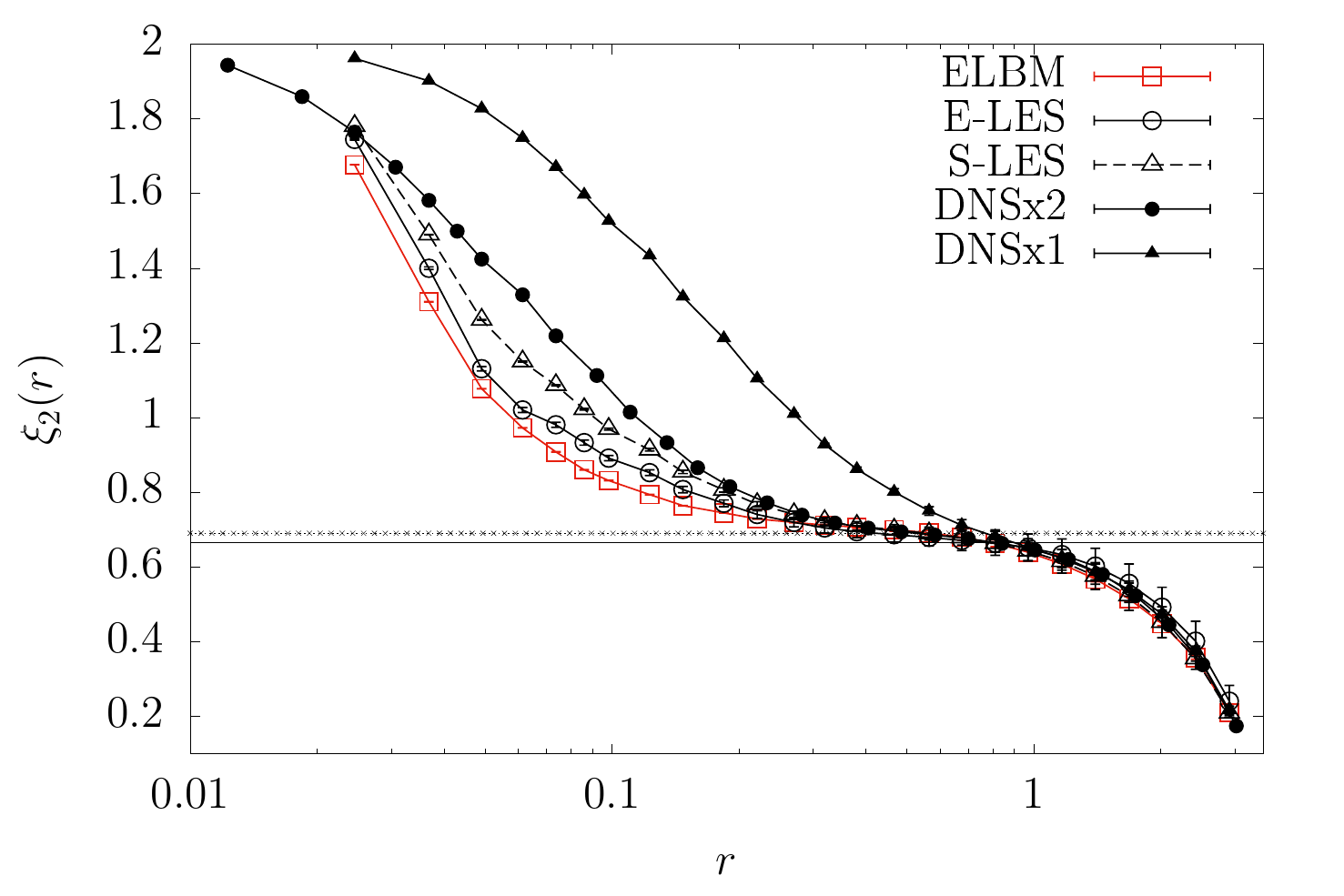}
  \includegraphics[width=8.5cm]{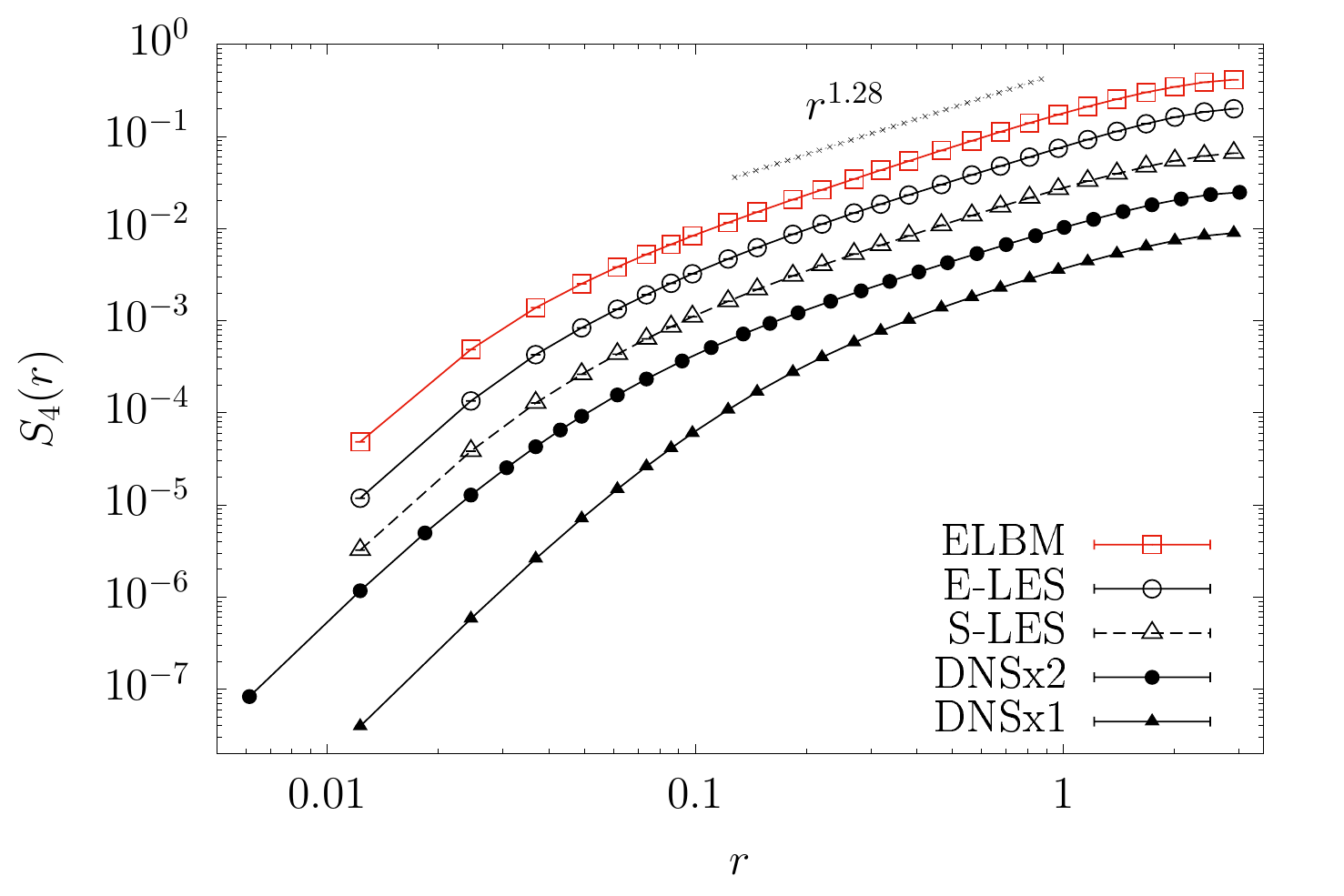}
  \includegraphics[width=8.5cm]{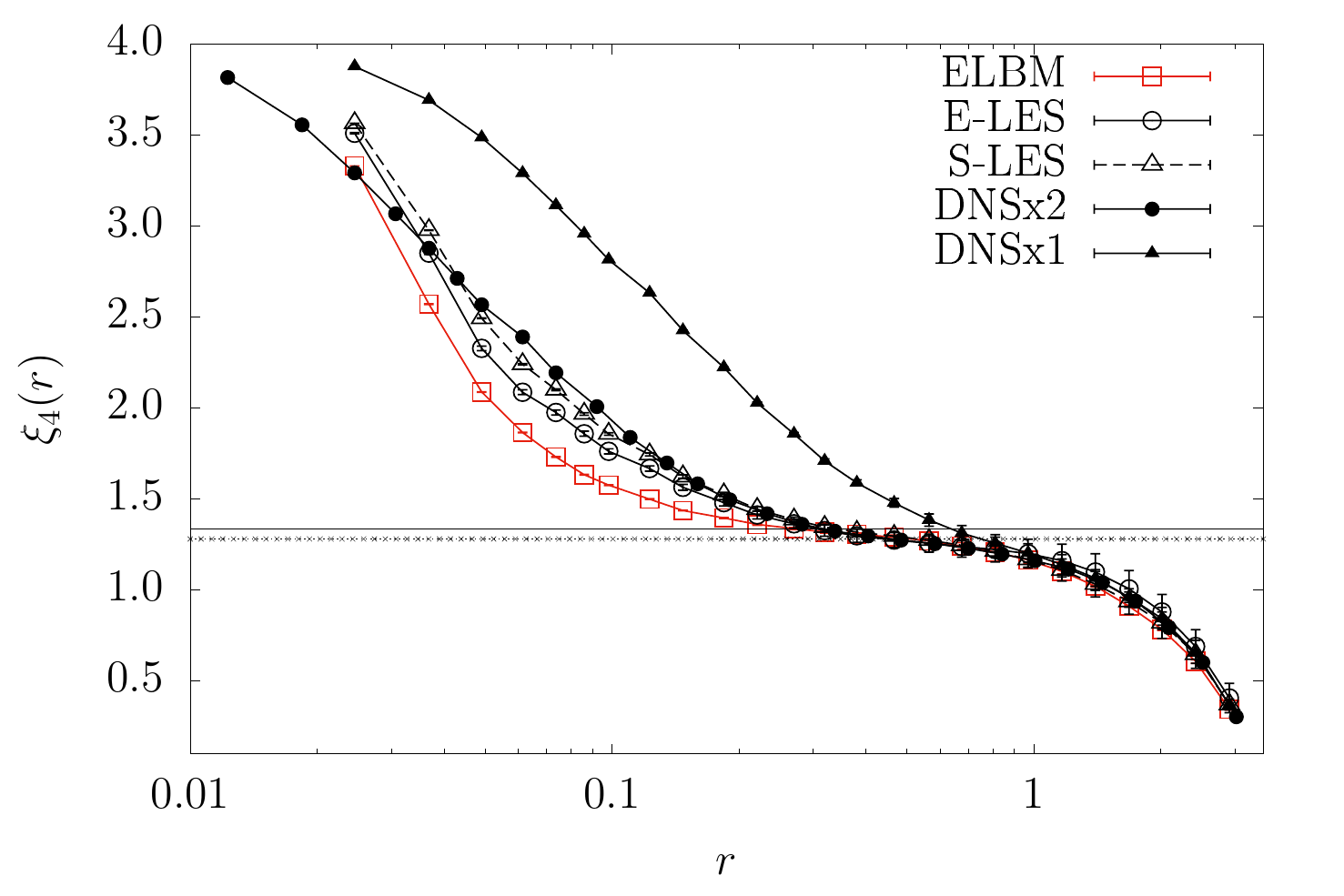}
\caption{Second-order longitudinal structure functions (left panels) and corresponding local slopes (right panels) for the conducted simulations at $512^3$ grid points, using ELBM (empty squares, red color), E-LES with entropic inspired model (empty circles), S-LES with Smagorinsky model (empty squares), DNSx1 at $512^3$ grid points (full triangles) and DNSx2 at $1024^3$ grid points (full circles). The straight line corresponds to the K41 prediction in the inertial range ($\xi_p=p/3$),  while the dashed line corresponds to the intermittent measure as reported from the literature~\cite{Gotoh2002,biferale2019self}.}
    \label{fig:sf_and_ls}
\end{figure*}

\begin{figure}
  \centering
  \includegraphics[width=0.5\textwidth]{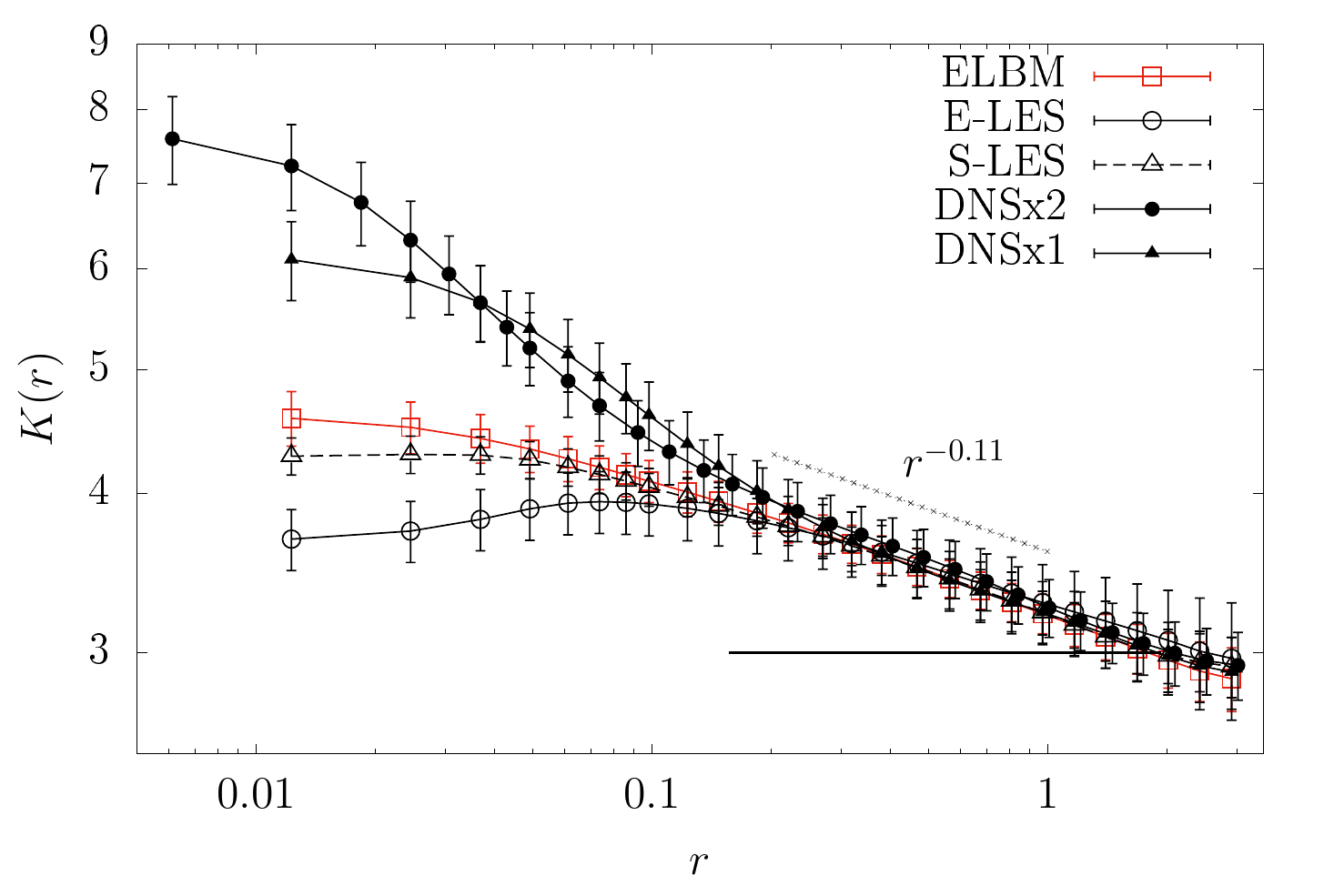}
  \caption{Kurtosis of the velocity increment for the simulations at $512^3$ grid points, using ELBM (empty squares, red color), E-LES with the entropic inspired model (empty circles), S-LES with Smagorinsky model (ampty triangles), DNSx1 at $512^3$ grid points (full triangles) and DNSx2 at resolution of $1024^3$ (full circles). The dashed horizontal line at 3 corresponds to the value of a Gaussian distribution.}
    \label{fig:kurtosis}
\end{figure}

\begin{figure}
  \centering
  \includegraphics[width=0.5\textwidth]{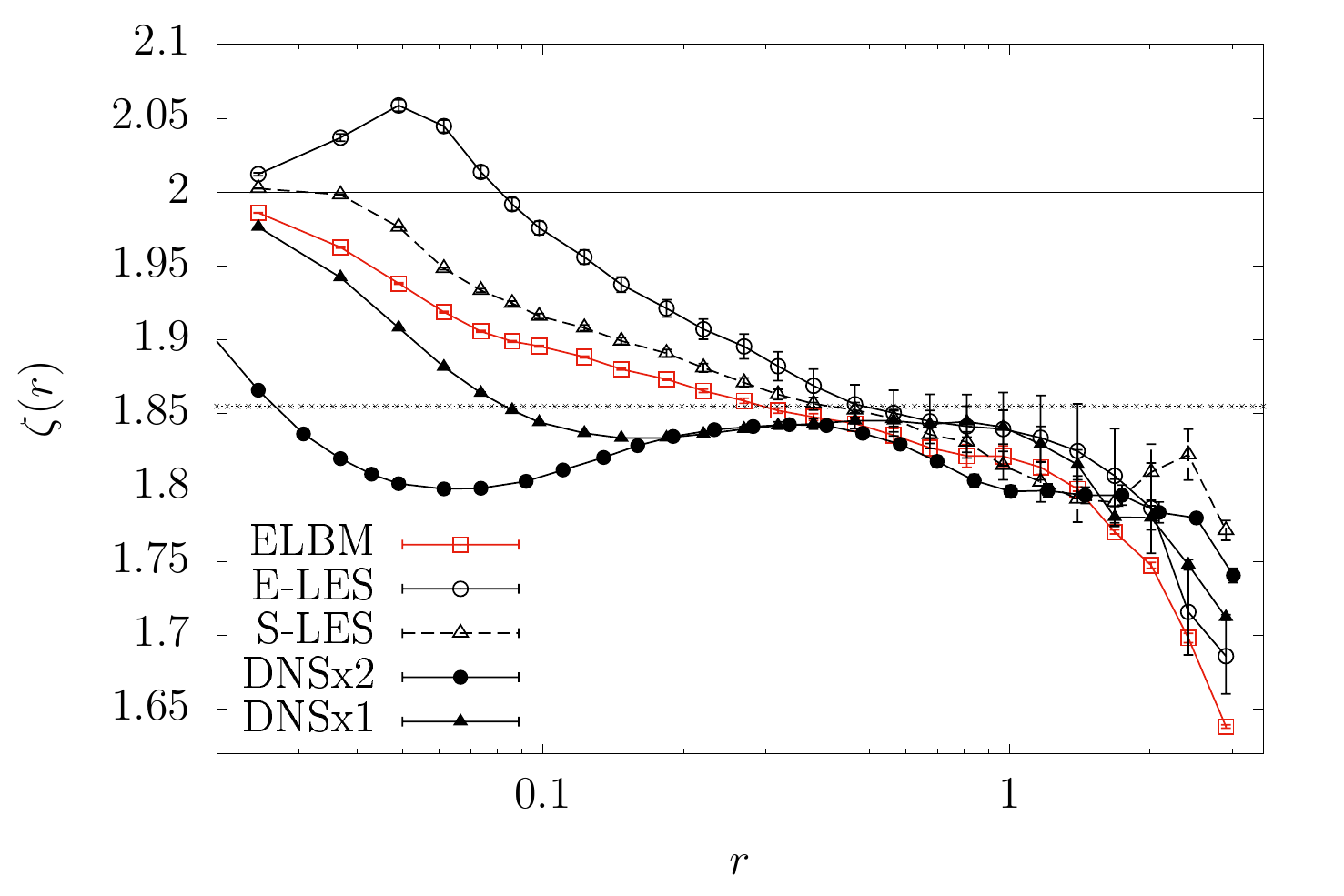}
  \caption{Extended Self-Similarity, $\zeta(r)$, for the simulations at $512^3$ grid points, using ELBM (empty squares, red color), the LES with entropic inspired model E-LES (empty circles), the LES with Smagorinsky model S-LES (empty triangles), the fully resolved DNSx1 (full triangles) and the fully resolved DNS at $1024^3$ grid points DNSx2 (full circles). The straight line corresponds to the K41 prediction in the inertial range equal to $2$, while the dashed line corresponds to the intermittent measure as reported from numerical~\cite{Gotoh2002} and experimental data~\cite{sinhuber2017dissipative}.}
  \label{fig:ess}
\end{figure}

In this last section we analyse the inertial range statistics by measuring the longitudinal velocity increments defined as $\delta_r u = (\bu(\bx+\br)-\bu(\bx))\cdot \br/r$. In this way we can quantify the effects produced by the different models at different scales, $r=|\br|$, hence, we can get an accurate estimation of the quality of the models in capturing the correct intermittent properties of the NSE. We study the scaling properties of the longitudinal structure functions (SF) defined as,
\RT{\begin{equation}
S_p(r) \equiv \langle [\delta_r u]^p\rangle 
\label{eq:SFlong}
\end{equation}
where the angular brackets indicate the ensemble average, that assuming spatio-temporal ergodicity can be evaluated averaging over space and time, $\langle (...) \rangle = \frac{1}{V}\frac{1}{T} \int_V \int_{t_0}^{t_0+T} (...)\, d\bx dt$.
In the limit of large Reynolds number, where $r$ can be taken arbitrarily small 
the structures function follows a powerlaw scaling behavior, $S_p(r)\sim r^{\xi_p}$ \cite{Frisch1995}, where a p-th order scaling exponent that, according to the phenomenological theory of Kolmogorov (K41)~\cite{Kolmogorov1941}, is  $\xi_p=p/3$.}
Nevertheless, both experimental and numerical studies have highlighted as bi-product of intermittency the presence of anomalous exponents in turbulent data, with important deviations from the K41 predicted values~\cite{Gotoh2002,sinhuber2017dissipative,benzi2010inertial,biferale2019self}. On the other hand, to get an accurate measurement of these exponents is extremely difficult. The reason is twofold, first it is required to have large scaling range (very well resolved simulations) and second it is simultaneously required to have large statistical ensemble. The first question we ask here is connected to the first of the aforementioned problems, namely whether those models are able or not to extend the length of inertial range of scales in our simulations. To answer this in the left column of Fig.~\ref{fig:sf_and_ls} we show the 2nd and 4th order structure functions measured from all modelled and fully resolved simulations. In the right column on the same figure we show the local scaling exponents, $$\xi_p(r) = \frac{d \, \log S_n(r)}{d \, \log (r)} \,,$$ measured form the structure functions shown on the left side. From these plots it is evident that all modelled simulations present an extension of the inertial range with respect to the same resolution, similar to the inertial range observed in the simulations with a number of grid points two times larger along each spatial direction, DNSx2.
Which means that the models allows to save a factor $8$ in terms of the number of degrees of freedom required in the simulations. Considering that also the time-step needs to be changed accordingly in the higher resolution DNS to resolve the smaller time-scales, it means that the modelled simulations are more than an order of magnitude cheaper than a DNS with the same inertial range. On the other hand, by measuring the local scaling exponents as reported on the right column of Fig.~\ref{fig:sf_and_ls}, we can see that the inertial range extension produced by the model is not as accurate as the fully resolved DNS. This is a problem if we want the model to correctly describe intermittency. Let us stress that the correction to the K41 prediction at the level of the second and fourth order exponents is very small, hence a model needs to be extremely accurate to correctly capture the intermittent scaling~\cite{biferale2019self}. In both right panels of Fig.~\ref{fig:sf_and_ls} we report with solid line the $\xi_p$ value of the K41 prediction, and with dotted lines the values measured from DNS as reported in~\cite{Gotoh2002,sinhuber2017dissipative}.
To highlight intermittency we can look at the ratio between SF at different orders. In particular, any systematic non-linear dependency of $\xi_p$ vs $p$, will introduce a scale-dependency in the Kurtosis, defined by the dimensionless ratio among fourth and second order SF,
\begin{equation}
K(r) =\frac{S_4(r)}{\left[S_2(r)\right]^2}.
 \label{eq:kurtosis}
\end{equation}
In Fig.~\ref{fig:kurtosis} we see that in all simulations, at large scales the increments are Gaussian ($K\sim 3$), while the Kurtosis quickly increases, decreasing the scale. This observation shows non-self similarity in the statistics of all data. 
It is interesting to observe that at this level the inertial range of scale observed in the DNSx2 simulation are well captured by all closures up to the dissipative scales $r \approx 0.1$ where deviations from the DNSs and models arise.
Going further, we measure the most refined quantity we can observe to quantify intermittency, namely the local scaling exponent in Extended Self-Similarity~\cite{Benzi1993},
\begin{equation}
    \zeta(r) = \frac{\xi_4}{\xi_2}.
    \label{eq:ESS}
\end{equation}
A linear K41 behavior would recover in the inertial range a plateau value of $\zeta$ equal to $2$. The correction, accounting for intermittency, measured in both experimental and DNS data gives the plateau for $\zeta$ at the value of $1.86$ ~\cite{Gotoh2002,sinhuber2017dissipative,biferale2019self}. As we can see in Fig.~\ref{fig:ess}, all models show deviations from the K41 self-similar prediction meaning that they all capture the non-self-similarity of the turbulent inertial range. However none of them is really accurate enough to extend the length of the plateau displayed, hence to improve the prediction obtainable from the DNS at the same resolution. Indeed, if we compare the modelled data with the fully resolved simulations we can see that former are not showing any flat plateau in the inertial range and we cannot estimate precisely the correction to K41 of the structure function scaling exponents.
It is interesting to point out that the models show a very similar accuracy up to this last analysis. This suggest the backscatter events of energy introduced by the entropic closures are not accurate enough to improve quantitatively the statistics of the Smagorinsky model. This results supports the observation that intermittency in turbulent flows comes as a result of highly non-trivial correlations among all degrees of freedom at different scales~\cite{buzzicotti2016phase,buzzicotti2016lagrangian,lanotte2015turbulence}. The observation that for all models we have a very similar inertial range statistics goes in agreement with the common property of these models to have the same scaling, proportional to the strain rate tensor.

\section{Conclusions}
\label{sec:6}

In this paper, we performed a quantitative assessment of the ELBM capabilities in the modelling of 3d homogeneous isotropic turbulent flows by comparing the inertial range statistics of ELBM data with the one of high resolution DNS of the NSE. We also compared the quality of ELBM with respect to the hydrodynamical Smagorinsky model, popular in the realm of LES. Furthermore, in this work we have proposed and investigated for the first time, a new hydrodynamical closure for LES simulation inspired from the macroscopic approximation of the ELBM model introduce by~\cite{Malaspinas2008}. We found that ELBM extends the length of the inertial range with respect to the fully resolved DNS, allowing to reduce the computational cost by an order of magnitude and at the same time preserving the simulation stability. Results showed that, in both the macroscopic and mesoscopic formulations, ELBM is able to reproduce an inertial range with a non self-similar dynamics. ELBM captures the correct deviations from the large scales Gaussian statistics as observed in the fully resolved DNS, with an accuracy comparable to the one produced by the Smagorinsky model.
From the measure of the structure functions scaling exponents in ESS, we have highlighted the limitations of these models to get with high accuracy the turbulence corrections to the Kolmogorov scaling. In this context we found that the modelled data are not producing the same inertial range plateau as observed in the fully resolved DNS and experiments. To conclude, we found that ELBM suffers in the modeling of extreme and rare intermittent fluctuations, while on the other hand it is very efficient in modeling the mean properties of 3d turbulence. Which makes ELBM a good candidate for the modeling of 3d turbulent flows in complex geometries.

\section*{Acknowledgement}

The authors thank Prof. Luca Biferale for inspiration and many useful discussions. This work was supported also by the European Unions Framework Programme for Research and Innovation Horizon 2020 (2014-2020) under the Marie Sk\l{}odowska-Curie grant [grant number 642069] and by the European Research Council under the ERC grant [grant number 339032]. The authors would like to thanks Prof. Dirk Pleiter as well as the Juelich Supercomputing Center for providing access to the JURON cluster.

\bibliographystyle{unsrt}
\bibliography{references}

\end{document}